\documentclass[a4paper,11pt]{article}
\pdfoutput=1 % if your are submitting a pdflatex (i.e. if you have
\usepackage[utf8]{inputenc}
\usepackage{jinstpub}
\usepackage{pgf}
\usepackage[normalem]{ulem}
\RequirePackage{lineno}
\usepackage{color}
\usepackage{upgreek}
\usepackage{tabularx}
\usepackage{booktabs}
\usepackage{sidecap}
\sidecaptionvpos{figure}{c}
\usepackage{caption,setspace}

\begin{document}

\title{GeMSE: a Low-Background Facility for Gamma-Spectrometry at Moderate Rock Overburden}

\affiliation[a]{Physikalisches Institut, Universität Freiburg, 79104 Freiburg, Germany}
\affiliation[b]{Institute of Geological Sciences, Universit\"at Bern, 3012 Bern, Switzerland}
\affiliation[c]{Natural History Museum Bern, 3005 Bern, Switzerland}
\affiliation[d]{Albert Einstein Center for Fundamental Physics, Universität  Bern, 3012 Bern, Switzerland}

\author[a]{Diego~Ram\'irez~Garc\'ia,}\emailAdd{diego.ramirez@physik.uni-freiburg.de}
\author[a]{Daniel~Baur,}
\author[a]{Jaron Grigat,}
\author[b,c]{Beda A.~Hofmann,}
\author[a]{Sebastian~Lindemann,}
\author[a]{Darryl Masson,}
\author[a,d]{Marc~Schumann,}
\author[d]{Moritz von Sivers,}
\author[a]{Francesco~Toschi}

\abstract{The GeMSE (Germanium Material and meteorite Screening Experiment) facility operates a low-background HPGe crystal in an underground laboratory with a moderate rock overburden of 620\,m.w.e.~in Switzerland. It has been optimized for continuous remote operation. A multi-layer passive shielding, a muon veto, and a boil-off nitrogen purge line inside the measurement cavity minimize the instrument's background rate, which decreased by 33\% to ($164\pm2$)\,counts/day (100\,--\,2700\,keV) after five years of underground operation. This agrees with the prediction based on the expected decay of short-lived isotopes. A fit to the known background components, modeled via a precise simulation of the detector, shows that the GeMSE background is now muon-dominated. We also present updates towards a more accurate detection efficiency calculation for the screened samples: the thickness of the crystal's outer dead-layer is precisely determined and the efficiency can now be easily calculated for any sample geometry. The advantage of this feature is showcased via the determination of the \textsuperscript{40}K content in the screening of a complex-shaped object: a banana.}

\keywords{ultra-low background, gamma spectrometry, HPGe, rare-event searches, banana}

%\linenumbers

\maketitle

%%%%%%%%%%%%%%%%%%%%%%%%%%%%%%%%%%%%%%%%%%%%%%%%%%%%%%%%%%%%%%

\section{Introduction}
\label{sec:introduction}

Low-background gamma-ray spectrometry based on high-purity germanium (HPGe) detectors is an indispensible tool for rare event search experiments in astroparticle physics, as it provides the means to select detector construction materials and components with the lowest possible contamination of radioactive impurities~\cite{Heusser:1995wd}. Other disciplines can also highly benefit from this technology. One example is geology and meteorite research, where short-lived (and thus hard to detect) isotopes are used to date the terrestrial age of a meteorite and can help to identify samples from the same fall~\cite{Leya:2009,Weber:2017}.

The GeMSE (Germanium Material and meteorite Screening Experiment) facility is installed in the Vue-des-Alpes tunnel in Switzerland~\cite{VdA}. It is mainly used for material screening for the XENON~\cite{Aprile:2017ilq,XENON:2020kmp,nT_radioassay} and DARWIN~\cite{DARWIN:2016hyl} dark matter projects and has a dedicated meteorite program, mostly centered around a large collection of meteorites collected in the Oman deserts~\cite{Rosen:2020aaa,Rosen:2021}. The facility was described in detail in~\cite{Sivers:2016yvm} and is thus only briefly summarized here: GeMSE is using a Canberra p-type coaxial HPGe crystal of 85\,mm diameter and 65\,mm height ($m\approx2.0$\,kg) installed in an ultra low-background U-style oxygen-free high thermal conductivity (OFHC)  copper cryostat. It is connected to a liquid nitrogen (LN\textsubscript{2}) bath via a coldfinger. GeMSE's massive shell-structured shield consists of (from inside to outside) 80\,mm of oxygen-free copper (Cu-OFE), 50\,mm of low-background lead ($^{210}$Pb concentration: ($7.2\pm0.5$)\,Bq/kg) and 150\,mm of standard lead. The large sample cavity is 240\,$\times$\,240\,mm$^2$ wide and 350\,mm high and accessed from the top via a massive sliding door. The entire shield is enclosed by a sealed glove box which is constantly purged with gaseous boil-off nitrogen (GN\textsubscript{2}), guided directly into the cavity. The depth of 620\,m.w.e.~(235\,m of rock overburden) of the laboratory in the Vue-des-Alpes tunnel reduces the muon flux to $8.3\times10^{-2}$\,m$^{-2}$s$^{-1}$. Two plastic scintillator panels of $1400 \times 1050$\,mm$^2$ are installed above and behind the glove box, read out by 3~photomultiplier tubes (PMTs) each, and serve as a muon veto.

Here we report on several notable updates and new features which significantly improved the performance of the instrument: like most HPGe screening facilities, GeMSE operates continuously. Over the last years it has been automated to an extent that visits to the laboratory are only necessary to refill the LN\textsubscript{2} inventory required to cool the crystal and to purge the shield (every $\sim$\,6\,weeks), as well as for sample changes. Full remote control further minimizes the workload for the operators. Detector operation is described in Section~\ref{sec:operation}. The background of the facility improved considerably over the last years, mainly due to the decay of cosmogenically-activated isotopes. The new background characterization presented in Section~\ref{sec:background} shows that the facility is now limited by muon-induced backgrounds which could be further reduced by rather straightforward upgrades to the muon veto. Finally, Section~\ref{sec:analysis} presents several upgrades to the analysis procedure: most importantly we are now able to calculate very accurate detection efficiencies also for  complex-shaped samples, which we demonstrate by screening a widely-known $^{40}$K souce -- a banana. The article closes with a summary and an outlook in Section~\ref{sec:summary}.

%%%%%%%%%%%%%%%%%%%%%%%%%%%%%%%%%%%%%%%%%%%%%%%%%%%%%%%%%%%%%%

\section{Detector Operation and Control}
\label{sec:operation}

The Vue-des-Alpes laboratory is a small underground space in the Swiss Jura mountains~\cite{VdA}. It does not provide any service besides electricity and network access, in particular there are no personnel on-site. This means that operators have to travel there by car for every action on the GeMSE detector. In order to simplify operation and to reduce travel to the absolute minimum, i.e., to refill the $\sim$\,360\,l of LN\textsubscript{2} storage in the laboratory and for sample changes, all standard operation procedures were automated and/or can be controlled remotely via the custom-designed slow control system Doberman (Detector OBsERving and Monitoring ApplicatioN)~\cite{Doberman}. 

\begin{figure}[]
    \centering
    {\includegraphics[width=\textwidth]{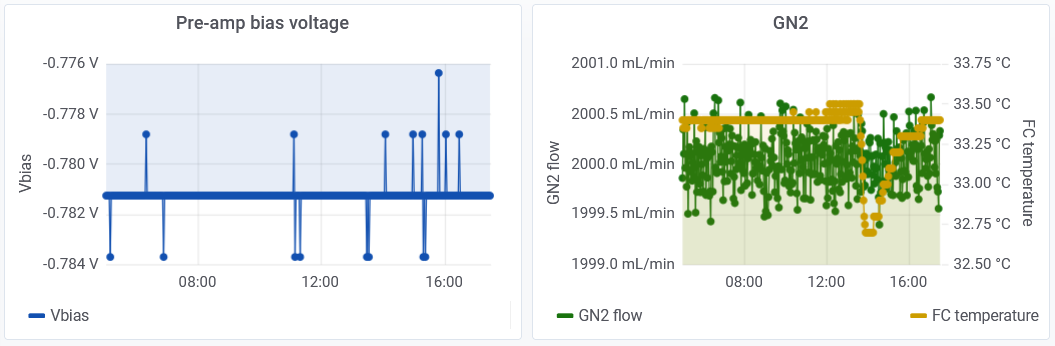}}
    {\includegraphics[width=\textwidth]{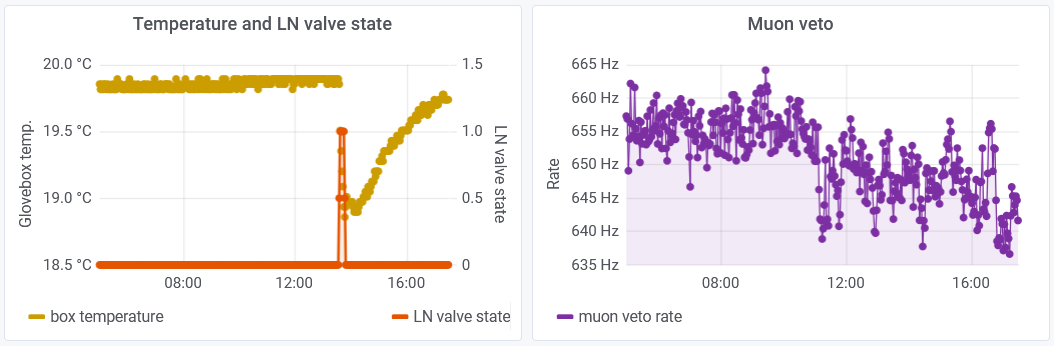}}
    {\includegraphics[width=\textwidth]{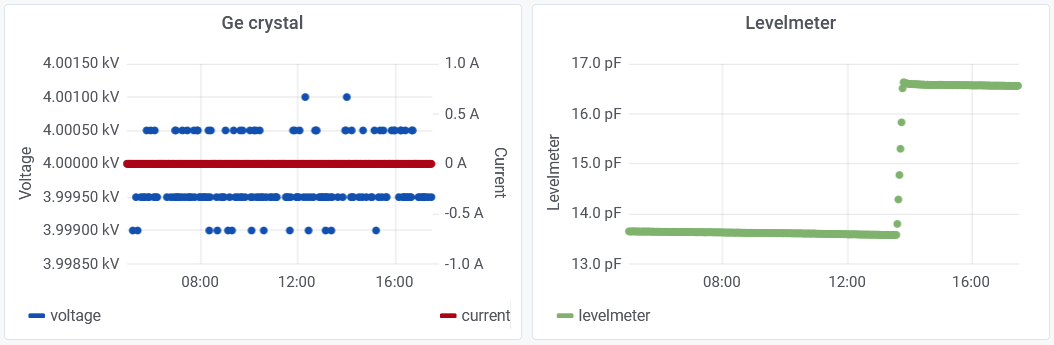}}
    \caption{Capture of the slow control monitor, showing a remote-controlled LN\textsubscript{2} refill during a stable operation period. From top to bottom and left to right, the panels show: HPGe detector pre-amplifier voltage; GN\textsubscript{2}-purge flow and the flow controller's temperature; temperature inside the glovebox and valve state of the LN\textsubscript{2} dewar connected to the detector coldfinger; muon veto trigger rate, with an average value of $\sim$\,650\,Hz; applied bias voltage to the HPGe crystal and current; and levelmeter capacitance, which can be translated into a liquid level inside the LN\textsubscript{2} dewar. The correlation between the open valve state (reading\,=\,1) and the levelmeter capacitance (liquid level) increase is apparent, as well as the temperature decrease inside the glovebox.}
    \label{fig:grafana}
\end{figure}

The slow control system was updated to enable operation of the facility for extended periods with only remote access. Slow control readings are handled asynchronously by a scheduling system to ensure independent, race-free readout of parameters from the same readout device. The data storage for the readings is realized with InfluxDB~\cite{influxdb} and a graphical web monitor is based on the Grafana dashboard~\cite{grafana}. A screen capture of the GeMSE slow control dashboard is shown in figure~\ref{fig:grafana}. The Doberman slow control system monitors all critical performance parameters, e.g., the level inside the LN\textsubscript{2} dewar connected to the detector coldfinger or the HPGe leakage current, and automatically sends SMS alarms in case readings fall out of pre-defined parameter ranges. It also allows for remote actions such as (un)biasing the HPGe crystal, refilling the LN\textsubscript{2} dewar connected to the crystal, and controlling the GN\textsubscript{2} purge flow. Further automation of the LN\textsubscript{2} handling can optimize the visit schedule and time of on-site detector operation: the slow control system can trigger a refill of the LN\textsubscript{2} dewar cooling the HPGe crystal from a storage vessel, using the readings of the capacitive sensor measuring the LN\textsubscript{2} level inside the dewar (see figure~\ref{fig:grafana}). The GeMSE data acquisition system can be also operated remotely but is not integrated in the slow control readings.

\begin{figure}[t]
    \centering
    {\includegraphics[width=\textwidth]{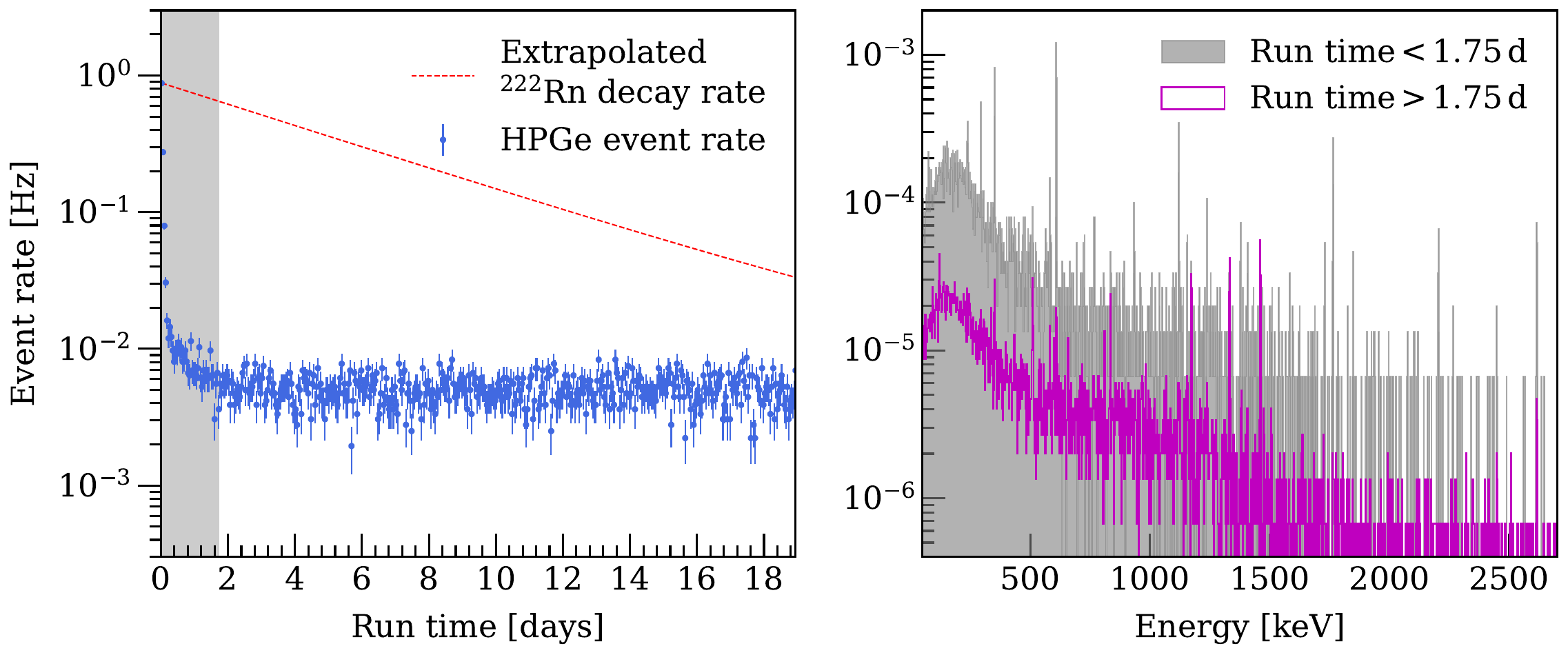}}
    \caption{Measured event rate of 10 Hamamatsu R11410-21 PMTs in the energy region between 50 and 2700\,keV. \textit{(Left)} Each data point corresponds to an hour. The dashed red line represents the expected rate without GN\textsubscript{2} purging, assuming the first measured data point is entirely attributed to \textsuperscript{222}Rn contamination. The non-flat part of the spectrum (1.75\,days, gray band) is discarded from the analysis. \textit{(Right)} The gray histogram corresponds to the first 1.75\,days and is populated by photopeaks from the \textsuperscript{222}Rn progeny (e.g., the 1765\,keV $\upgamma$-line from \textsuperscript{214}Bi). The magenta histogram with significantly fewer entries corresponds to the remaining $\sim$\,17\,days of measurement. Peaks from, e.g., \textsuperscript{60}Co (1172\,keV, 1332\,keV) and \textsuperscript{40}K (1461\,keV) are clearly visible as they do not stem from the air contamination from the first part of the acquisition.\label{fig:gemse_rate}}
\end{figure}\label{page:ge_rate_plot}

A typical sample is measured from a few days, for freshly fallen meteorites, up to 6\,weeks, for ultra-pure material candidates for rare-event search experiments, depending on its size and radioactive contamination. A 14-bit digital multichannel analyzer (CAEN DT5781A MCA), with real-time digital pulse processing~\cite{MCA_caen} records the signals from the HPGe detector. It is operated via the manufacturer's software in time-stamped list mode~\cite{MC2_caen} which enables the recording of pulse height and time of every event at 10-ns resolution. The latter is especially useful for data selection: periods with higher rate in the detector cavity, right after a sample exchange, or failures in the GN\textsubscript{2} purge or the muon veto, can be removed from data before sample analysis. The effect of purging on the decrease of the effective acquisition is shown in figure~\ref{fig:gemse_rate} where 10~low-background Hamamatsu R11410-21 PMTs used for the XENONnT experiment were screened~\cite{1t_pmts, nT_radioassay}.

The signals of the muon veto PMTs are amplified before being fed into a discriminator and then a coincidence trigger. A gate generator ensures a fixed width of the trigger signal which is then fed into the MCA as a veto. A muon veto occurs when at least one of the scintillator panels detects a hit which involves two of the three PMTs in coincidence. This signal vetoes the MCA for 10\,$\upmu$s, inducing an acquisition dead time of about 0.6\%.

%%%%%%%%%%%%%%%%%%%%%%%%%%%%%%%%%%%%%%%%%%%%%%%%%%%%%%%%%%%%%%

\section{Background Characterization}
\label{sec:background}

An integral background rate of ($246\pm2$)\,counts/day in the region of interest 100\,--\,2700\,keV was measured in 2016, right after the detector installation underground~\cite{Sivers:2016yvm}. Cosmic neutron-induced spallation is responsible for the production of most of the short-lived cosmogenic radionuclides in the detector and shield materials at the earth's surface, while this production mechanism can be already considered negligible at depths of a few tens of m.w.e.~\cite{Heusser:1995wd,Cebri_n_2017,Cebri_n_2020,ZHANG201662,Baudis_2015}. At a moderate rock overburden of 620 m.w.e.~these isotopes are thus expected to decay away, further reducing the already ultra-low initial background rate.

\begin{figure}[]%h
    \centering
    {\includegraphics[width=\textwidth]{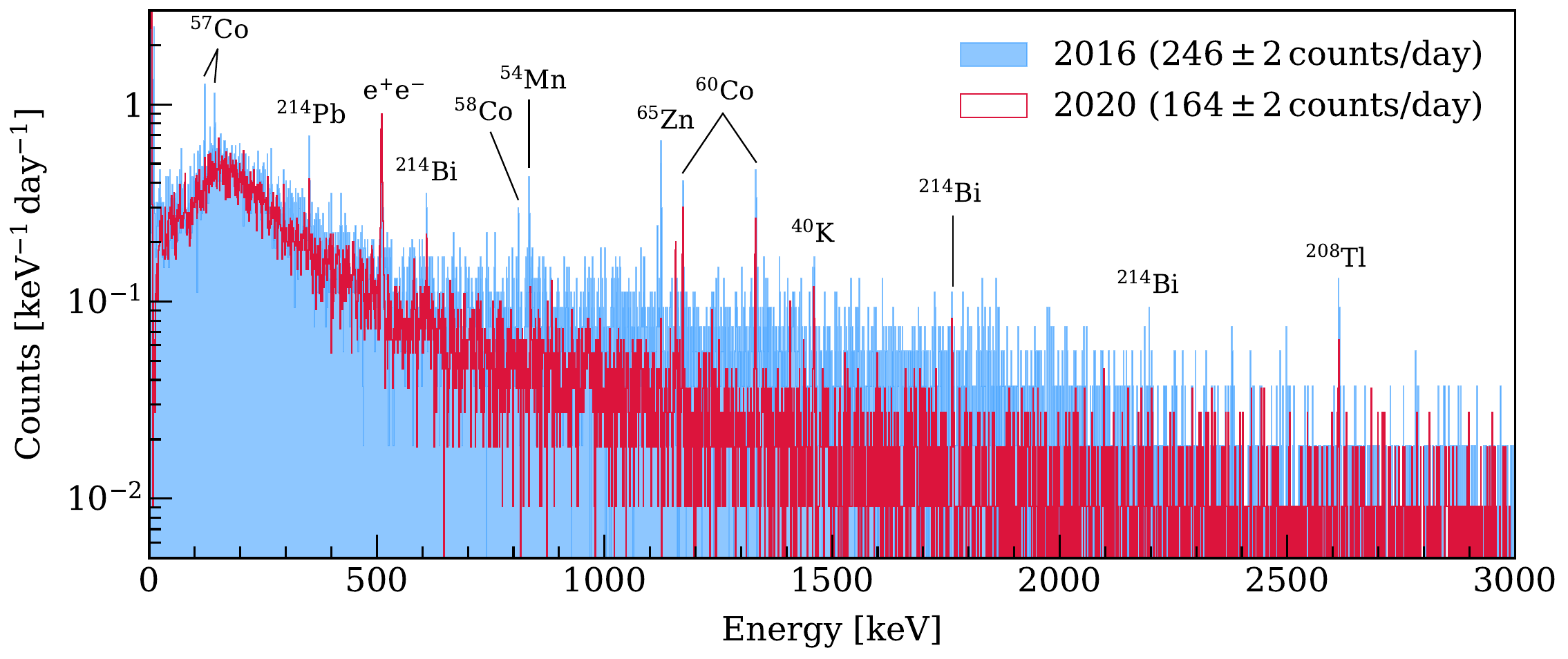}}
    \caption{\label{fig:gemse_bkg}GeMSE background spectrum during the early detector operation (2016~\cite{Sivers:2016yvm}, blue) and now (2020, red). Some $\upgamma$-ray lines are labeled, as well as the e\textsuperscript{$+$}e\textsuperscript{$-$} pair production line at 511\,keV. A rate reduction of $\sim$\,33\% occurred after about four years of underground operation due to the decay of the short-lived cosmogenic isotopes. By operating GeMSE underground, the activation rate is strongly suppressed, therefore the background rate decreases with the half-life of these isotopes (mainly \textsuperscript{56-58}Co, \textsuperscript{54}Mn, \textsuperscript{65}Zn and \textsuperscript{68}Ge).}
\end{figure}

\paragraph{Background Description} A new combined background run of 109.7\,days was acquired in 2020. Figure~\ref{fig:gemse_bkg} shows the comparison with the background from 2016 with some $\gamma$-lines of interest labeled. The new background rate is ($164\pm2$)\,counts/day or ($82\pm1$)\,counts/(day\,$\cdot\,$kg) in the 100\,--\,2700\,keV region (($91\pm1$)\,counts/(day\,$\cdot\,$kg) in 40\,--\,2700\,keV). This number is in excellent agreement with the predicted rate of $\sim$\,80\,counts/(day\,$\cdot\,$kg) in~\cite{Sivers:2016yvm}, calculated by applying the half-life correction of the activated isotopes to the 2016 background spectrum. 
This background level places GeMSE among the most sensitive HPGe detectors available~\cite{Laubenstein:2020rbe}, even though it is operated only at a moderate depth.

\begin{figure}[]
    \centering
    {\includegraphics[width=\textwidth]{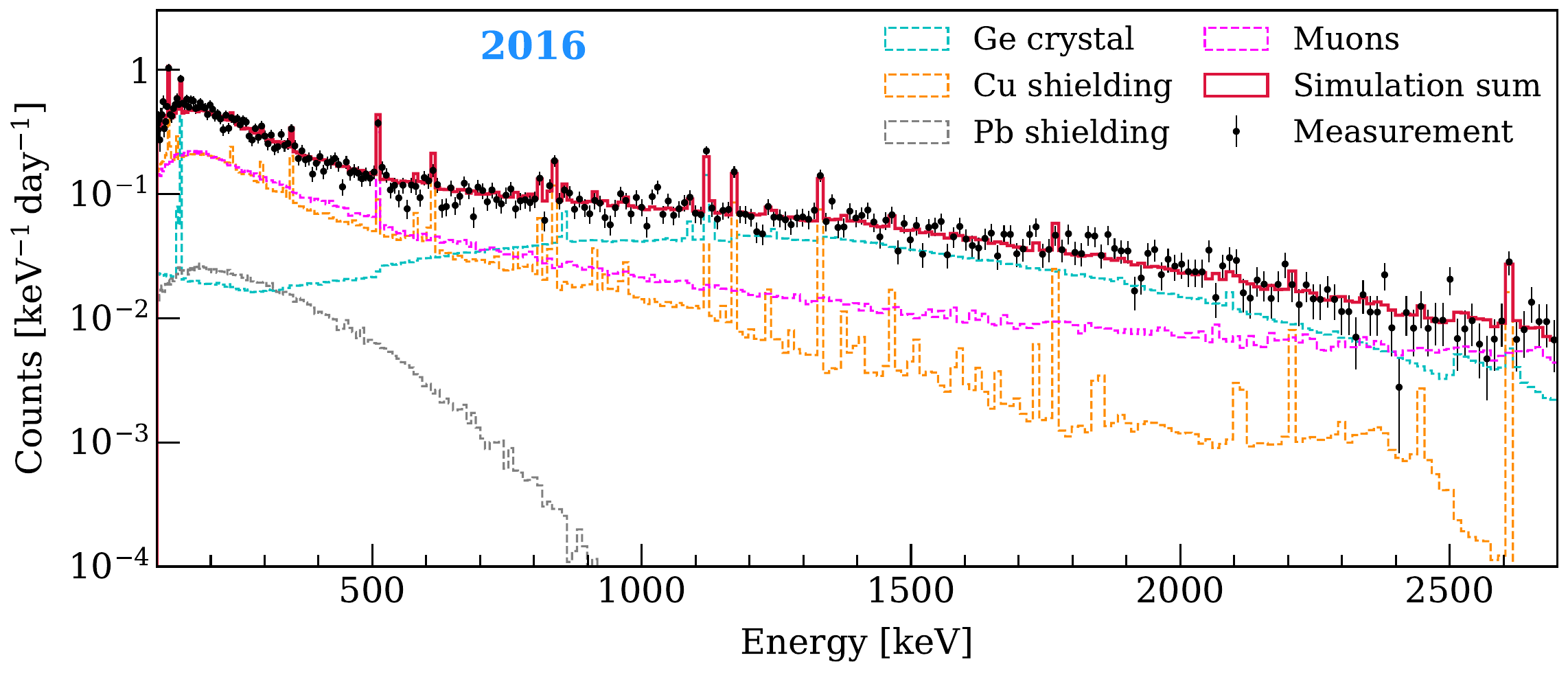}}\\
    \vspace{0.3cm}
    {\includegraphics[width=\textwidth]{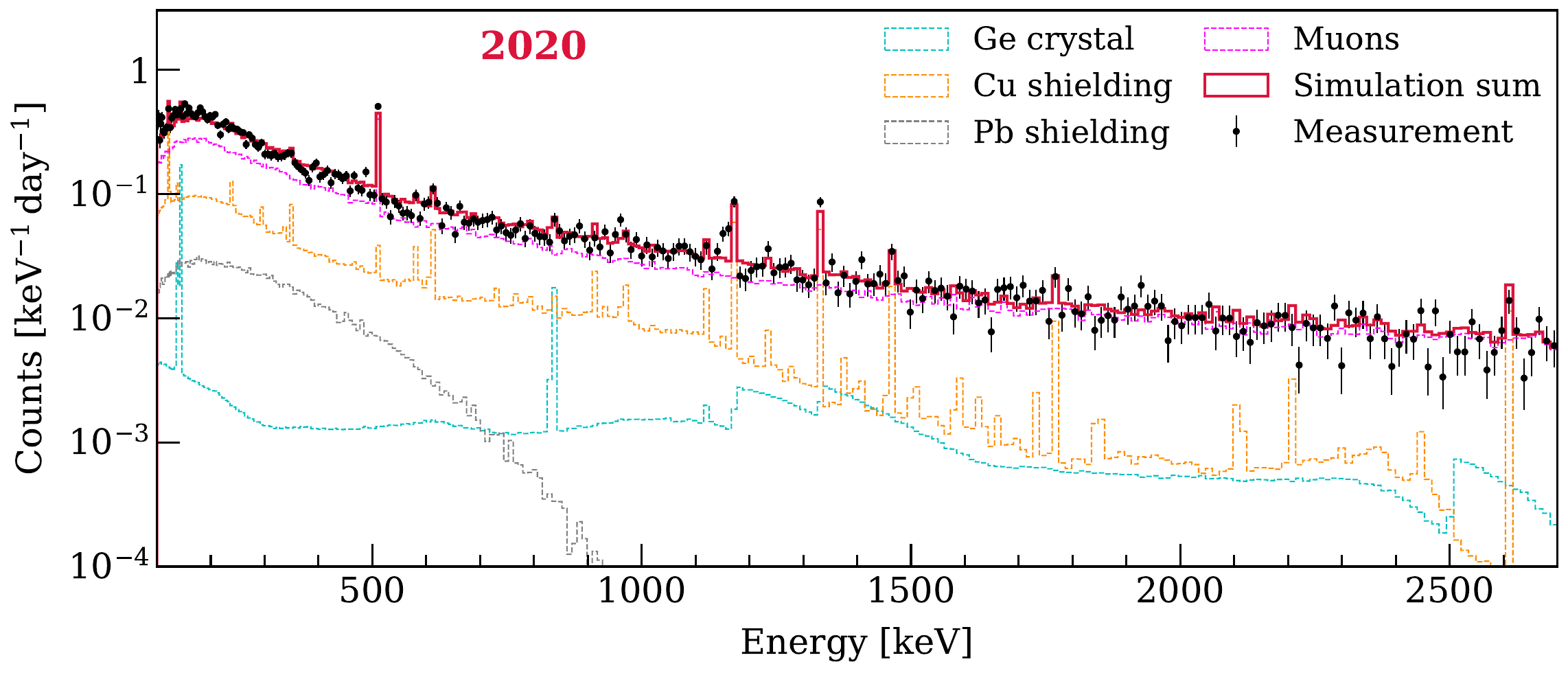}}
    \caption{Best fit (solid red line) to the measured GeMSE background spectrum (black points) in 2016 \textit{(top)} and 2020 \textit{(bottom)}. The  contributions from the Ge crystal, Cu and Pb shield, and muons are shown as dashed lines. The deactivation of the Ge crystal (cyan), discussed in the text, is mainly due to the suppression of the \textsuperscript{68}Ga $\upbeta$-spectrum. The fit for the 511\,keV e\textsuperscript{$+$}e\textsuperscript{$-$} peak shows that the contribution attributed to muon-induced events has slightly increased compared to the background measurement from 2016. The fit slightly underpredicts the measured data in the  \mbox{$\sim$\,100\,--\,200\,keV} region, whis is expected as the measured Compton continuum in this range also includes not simulated residual background sources. \label{fig:gemse_bkg_fit}}
\end{figure}

Figure~\ref{fig:gemse_bkg} reveals a significant reduction of the background from short-lived cosmogenically-activated isotopes which were present during early detector operation. The most important differences between the background spectra from 2016 and 2020 are:
\begin{itemize}
    \item the suppressed continuous $\upbeta$-spectrum of $^{68}$Ga ($Q=2921.1$\,keV), a short-lived daughter of $^{68}$Ge ($T$\textsubscript{1/2}\,=\,270.8\,d), which leads to the visible rate reduction in the central region of the plot (maximum around $\sim$1300\,keV, shown in detail in figure~\ref{fig:gemse_bkg_fit});
    \item the activity reduction for the characteristic $^{65}$Zn ($T$\textsubscript{1/2}\,=\,244\,d) peak at 1115.5\,keV; % following an electron capture decay.%This energy does not correspond to a transition in particular, but is the result of the sum of the Auger electrons following the \textsuperscript{65}Zn electron capture decay and the 1115.5\,keV $\upgamma$-ray emission from the excited daughter isotope.
    \item the activity reduction for the $^{54}$Mn ($T$\textsubscript{1/2}\,=\,312.2\,d) $\upgamma$-line at 834.8\,keV;%, following a decay via electron capture with branching ratio close to 100\%.
    \item the activity reduction for some prominent $\upgamma$-peaks from the short-lived cobalt isotopes: the 122.1\,keV and 136.5\,keV $\upgamma$-lines from ${}^{57}$Co (T${}_{1/2}$=271.7 d), following the electron capture decay into $^{57}$Fe, and the $\upgamma$-line of $^{58}$Co ($T$\textsubscript{1/2}\,=\,70.9\,d) at 810.8\,keV; % with BR\,$>$\,99\%.
    \item the current rate of the 511\,keV line from e\textsuperscript{$+$}-e\textsuperscript{$-$} annihilation is slightly larger than in 2016. Positrons arise from the interaction of muons in the detector surroundings, which points to a potentially deteriorated muon veto tagging efficiency. This is further discussed below after evaluating the rate for each background contribution to the overall spectrum.
\end{itemize}

\paragraph{Background Model and Analysis} The background is analyzed by matching the measured counting rates with the spectra from the different predicted background contributors. These are obtained by means of detailed Monte Carlo simulations using the Geant4 toolkit~\cite{Agostinelli:2003geant4, ALLISON2016186}. The GeMSE simulations framework is based on Geant4 version~10.7 and is publicly available~\cite{gemse_mc}. The flux of cosmic-ray muons is simulated using the \texttt{Shielding} Geant4 physics list, which accounts for spallation processes and the propagation of secondaries. Entries in the crystal are vetoed for 10\,$\upmu$s following events producing an energy deposit $>$1\,\,MeV in one of the muon veto panels. The measured muon flux of $(8.3\pm0.2) \times 10^{-2}$\,m$^{-2}$\,s$^{-1}$ is used as a prior for the fit for this component.

Besides the muons, the model includes the combined contributions from radioactive isotopes in the HPGe crystal as well as in the Cu and Pb shielding. The background contribution from the residual \textsuperscript{222}Rn in the sample cavity ($\sim$\,0.3\,mBq/m\textsuperscript{3}) and the crystal's Cu cryostat ($\sim$\,0.8\,mBq/kg) are negligible~\cite{Sivers:2016yvm} and therefore not included in the final fit. However, some degeneracy for the background from the Cu shielding and the detector cryostat is expected given their similar spectral shape. The combined rate for the \textsuperscript{56}Co and \textsuperscript{58}Co isotopes in the Ge~crystal and Cu~shielding was derived as an upper limit in the 2016 analysis. Since $\sim$\,20 half-lives for these isotopes have elapsed between the two background measurements, and under the reasonable assumption of no new cosmogenic activation underground at 620\,m.w.e., these isotopes are also excluded from the new fit.

The simulated spectra are fitted to the new background measurement and Bayesian inference is used to determine the detected values/upper limits of the isotopic activities~\cite{Sivers:2016yvm}. Figure~\ref{fig:gemse_bkg_fit} shows the best fits to the newly measured background spectrum from 2020, as well as the one from 2016 for comparison. A variable bin size proportional to the detector resolution function is used. The p-value of the fit is calculated according to~\cite{PhysRevD.83.012004} and is~0.01. This indicates a good match with the background model, considering the  wide fit range and the fact that some underprediction is expected especially at low energies, where the measurement contains contributions from the Compton spectra of all subdominant background sources ignored in the model.

The background contributions induced by the Cu shield (orange) and the HPGe crystal itself (cyan) have significantly decreased as a result of the decay of cosmogenically-activated isotopes. This effect is most striking for the HPGe crystal, which dominated the background in 2016 with the \textsuperscript{68}Ga $\upbeta$-spectrum. Now the expected features from other isotopes are also distinguishable, e.g., the \textsuperscript{60}Co $\upgamma$-rays at 1173\,keV and 1332\,keV and their sum peak at 2505\,keV\footnote{The fin-like asymmetric shape and the relative height of the sum peak have geometrical reasons: since the \textsuperscript{60}Co$\to$\textsuperscript{60}Ni$^{*}$ decay happens within the HPGe crystal, the $\upbeta^{-}$ emission is recorded in most of the cases in coincidence with the gamma lines resulting from the \textsuperscript{60}Ni$^{*}$ de-excitation.}. The \textsuperscript{210}Pb activity extracted from the fit agrees with the value from external measurements of lead used in the GeMSE shield.

The "effective" (i.e., not vetoed) muon flux derived from the fit has increased by~20\% compared to 2016, probably due to a worsened tagging efficiency of the setup. The fit successfully models this component, although the model lacks precise knowledge about the features of muon identification in the laboratory such as the exact rock composition/profile and the validity of the energy threshold for the simulated muon veto tagging efficiency. %An independent efficiency measurement for the muon veto panels must be performed to better constrain this fit parameter. 
The resulting activity of the considered isotopes is summarized in table~\ref{tab:bkg_gemse} and compared to these derived from the 2016 background spectrum~\cite{Sivers:2016yvm}.

\begin{table}[t]
\caption{Radioactive contaminants contributing to the GeMSE background extracted from the fit of the simulated components to the measured spectrum. A detection is quoted with uncertainties (68\% confidence interval) while upper limits are defined by the 95\% confidence interval. The values from 2016 are taken from~\cite{Sivers:2016yvm}. The muon flux quoted here corresponds to an "effective" flux passing the muon veto.}
\resizebox{\textwidth}{!}{%
\centering
%\small
\begin{tabular}{lcccccccccc}
\toprule
&  \multicolumn{10}{c}{Specific isotope activity [$\upmu$Bq/kg]} \\[-0.15cm]
Source & \cline{1-10}
& $^{65}$Zn & $^{68}$Ge & $^{60}$Co & $^{58}$Co & $^{57}$Co & $^{56}$Co & $^{54}$Mn & $^{238}$U & $^{232}$Th & $^{40}$K \\
\midrule
\textcolor{gray}{Ge crystal (2016)} & \textcolor{gray}{77${+15 \atop -13}$} & \textcolor{gray}{313${+17 \atop -23}$} & \textcolor{gray}{$<118$} & \textcolor{gray}{$<3.5$} & \textcolor{gray}{11${+2 \atop -1}$} & \textcolor{gray}{$<43$} & \textcolor{gray}{$<23$}& \textcolor{gray}{--} & \textcolor{gray}{--} & \textcolor{gray}{--} \\[0.14cm]
Ge crystal (2020) & $<5.7$ & $<15.1$ & $<26.4$ & -- & 3.6${+1.3 \atop -1.1}$ & -- & $<5.4$ & -- & -- & --\\[0.22cm]
\textcolor{gray}{Cu shield (2016)} & \textcolor{gray}{--} & \textcolor{gray}{--} & \textcolor{gray}{43${+7 \atop -5}$} & \textcolor{gray}{$<26$} & \textcolor{gray}{374${+48 \atop -39}$} & \textcolor{gray}{$<7.6$} & \textcolor{gray}{$<56$} & \textcolor{gray}{104${+13 \atop -15}$} & \textcolor{gray}{42${+9 \atop -10}$} & \textcolor{gray}{$<196$}\\[0.13cm]
Cu shield (2020) & -- & -- & 35${+3 \atop -3}$ & -- & 121${+23 \atop -24}$ & -- & $<11.5$ & 53${+7 \atop -7}$ & 30${+5 \atop -6}$ & 103${+36 \atop -29}$\\[0.08cm]
\midrule
\textcolor{gray}{Pb shield (2016)} & \multicolumn{10}{l}{\textcolor{gray}{$^{210}$Pb activity $<8.2$\,Bq/kg}}\\[0.14cm]
Pb shield (2020) & \multicolumn{10}{l}{\hphantom{$^{210}$Pb activity }$=7.4{+0.5 \atop -0.5}$\,Bq/kg}\\[0.22cm]
\textcolor{gray}{Muons (2016)} & \multicolumn{10}{l}{\textcolor{gray}{Flux = $1.19{+0.07 \atop -0.09} \times 10^{-5}$\,cm$^{-2}$\,s$^{-1}$}}\\[0.14cm]
Muons (2020) & \multicolumn{10}{l}{\hphantom{Flux = }$1.44{+0.04 \atop -0.05} \times 10^{-5}$\,cm$^{-2}$\,s$^{-1}$} \\
\bottomrule
\end{tabular}}\label{tab:bkg_gemse}
\end{table}

\paragraph{Muon Veto Efficiency}

By integrating the fit spectra in figure~\ref{fig:gemse_bkg_fit} we obtain that about~70\% of the current background in the GeMSE detector is caused by muon-induced events, even after the reduction by the muon veto. This motivates studies  to upgrade the muon veto by enlarging the geometrical coverage of the cavity with additional plastic scintillator panels.

In order to better understand the current muon veto capabilities of GeMSE, as well as to confirm the spectral shape of this component used in the background model, a measurement simultaneously recording interactions in the Ge crystal and the muon veto was performed. Contrary to the default settings, in which reading the signals from the HPGe detector is directly vetoed after a muon veto trigger, in this measurement the (logic) muon veto trigger signal was read out by using the second channel of the MCA recording the time stamps and pulse heights of the signals from the HPGe crystal. Both MCA channels feature the same clock, i.e., the time stamps from both channels can be directly compared; it was verified that muon veto signals occurring in coincidence with signals from the crystal were not delayed with respect to the latter by their acquisition electronics.

\begin{figure}[]%h
    \centering
    %\vspace{1cm}
    {\includegraphics[width=\textwidth]{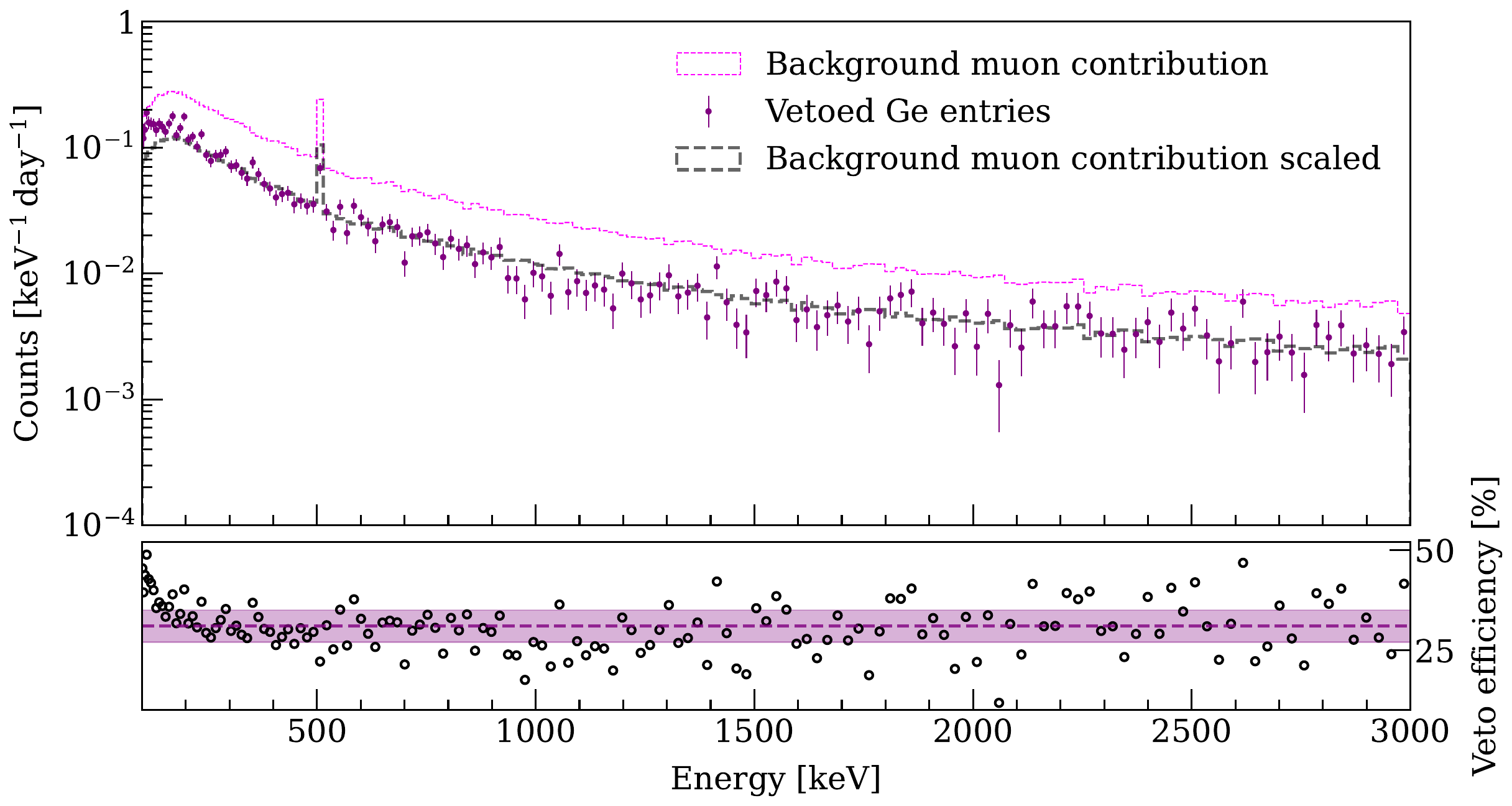}}
    \caption{Measured HPGe detector signals in coincidence with the muon veto detector (purple dots). Their spectral shape matches that of the predicted  muon-induced background extracted from the multi-component fit to the background data (magenta histogram), i.e., remaining after the active muon veto. The comparison of vetoed (data points) and total (data points plus histogram) muon events yields a muon veto tagging efficiency ($31\pm4$)\%, shown in the bottom panel with the 1$\sigma$ uncertainty. The increased efficiency for energies $<200$\,keV can be explained by the accidental pairing of events in this region (happening at higher rate) with fake muon veto triggers produced by readout noise in the scintillator panels. The gray histogram shows the fitted background contribution (magenta histogram) scaled according to this tagging efficiency.\label{fig:gemse_muons}}
\end{figure}

To emulate the finite veto window of the standard acquisition mode, the muon veto data were searched for entries within the 10\,$\upmu$s interval preceding every recorded HPGe detector signal. This allowed selecting the Ge events with a coincident muon veto signal which are otherwise discarded. The measured spectrum of muon-induced Ge events is shown in figure~\ref{fig:gemse_muons} (purple data points) together with the simulated muon veto spectrum fitted to the GeMSE background (magenta histogram) from figure~\ref{fig:gemse_bkg_fit}. Their spectral shapes agree very well. It must be noted that the data points represent entries recorded by the HPGe detector which were identified by the muon veto, while the histogram from the fit corresponds to the remaining, non-tagged muon-induced background. The ratio of tagged and total muon events can be used to estimate the muon veto tagging efficiency, shown in the bottom panel. The average efficiency in the range 100\,--\,2700\,keV (dashed purple line) is ($31\pm4$)\%. The agreement of the spectral shapes of the simulated and the recorded muon-induced data validates the Monte Carlo prediction for this background component.

%%%%%%%%%%%%%%%%%%%%%%%%%%%%%%%%%%%%%%%%%%%%%%%%%%%%%%%%%%%%%%

\section{Improvements in Sample Analysis}
\label{sec:analysis}

The analysis of the sample spectra measured with GeMSE generally follows the procedure described in~\cite{Sivers:2016yvm}, but features several upgrades to the simulations framework used to estimate the detection efficiency as well as the MCA software and firmware. The utilized software framework is available at~\cite{gemse_container} as a Docker container~\cite{docker} and includes the Bayesian Analysis Toolkit (BAT)~\cite{bat_ref} required to evaluate activities for ultra low-background counting rates.

In order to obtain the activity for a given radioisotope, corrections need to be applied for the branching ratio of the observed $\upgamma$-lines as well as for the detection efficiency of the HPGe crystal. This efficiency depends on the energy and the geometrical placement of the sample inside the measurement cavity with respect to the HPGe crystal and shield. It is therefore necessary to accurately model the combined detector-sample geometry in the GeMSE simulations, also to ensure that self-shielding effects are properly accounted for.

\paragraph{Estimating the Dead Layer Thickness}

The 1 mm-thick OFHC copper cryostat prevents low-energy X-rays from penetrating the Ge crystal. In addition to the cryostat shielding, special care must be taken to properly model the dead layer of the crystal itself. This dead layer refers to the n$^+$-contact consisting of Li~atoms diffused into the Ge; it behaves as an insensitive region which only absorbs radiation. Its thickness is smallest just after the production of the HPGe crystal and gradually increases with time as the Li~atoms diffuse into the crystal, a process greatly inhibited by cryogenic conditions~\cite{LABORIE200057, QUANGHUY2010390, huy}. An increase of the inactive dead layer reduces the remaining sensitive volume and leads to a decrease in the detector efficiency. The p$^+$-contact, implanted in the inner region of the crystal, has a thickness of $\sim$\,0.3\,$\upmu$m and thus does not significantly affect the efficiency~\cite{proceedings_p_n}.

A dead layer thickness of ($0.67\pm0.01$)\,mm was inferred in 2016 by using a certified multi-gamma calibration source~\cite{Sivers:2016yvm}, which was then implemented in the simulated model. An increase in the dead layer thickness was expected due to the few thermal cycles that the detector underwent in the past years, including a longer three-month shutdown during the start of the COVID-19 pandemic in Europe. A new determination of the dead layer was conducted using the peak-ratio approach~\cite{Jiang_2016, loan_ge}: a spectrum from a calibrated \textsuperscript{133}Ba point-source located at a precisely known distance from the top of the detector cryostat (138\,mm) was recorded, and the geometry of this calibration source and its holder was implemented in the Geant4 detector model. The ratio of the signal strengths of the 81\,keV and 356\,keV full-absorption peaks was measured and compared to the ratio obtained from simulations for different thicknesses of the passive layer.

\begin{figure}[]
    \centering
    {\includegraphics[width=\textwidth]{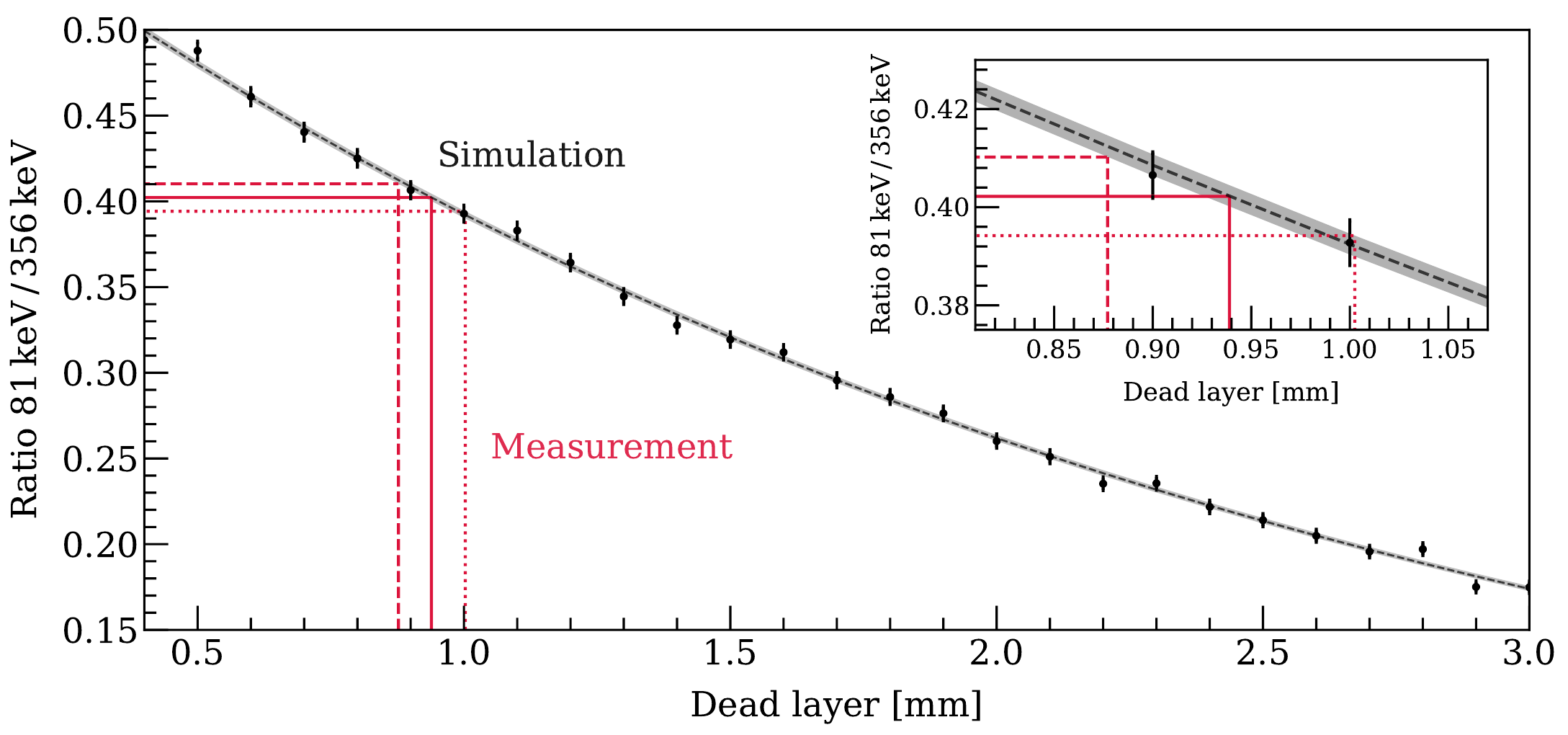}}
    \caption{The dead layer thickness is obtained using the ratio of the signal strengths of the 81\,keV and 356\,keV full absorption peaks of a calibrated \textsuperscript{133}Ba point-source placed 138\,mm above the cryostat endcap. Every data point corresponds to one simulation ($4\times10^6$ primaries) for a given dead layer. The exponential fit to this data and the 1$\sigma$ uncertainty are represented by the dashed line and the gray band, respectively. The experimentally measured ratio and its uncertainties are given by the red solid and dashed/dotted lines, respectively. The obtained dead layer thickness is ($0.94\pm0.06$)\,mm.}
    \label{fig:ge_peak_ratio}
\end{figure}

Figure~\ref{fig:ge_peak_ratio} shows the simulated signal strength ratios, which can be well described by an exponential function (dashed black line), as well as the measured ratio (red solid line) with uncertainties. By comparing the experimental ratio and the simulated line, and repeating the process for the measured uncertainty and the 1$\sigma$ band of the exponential fit, the detector's effective dead layer thickness of ($0.94\pm0.06$)\,mm is obtained. This corresponds to an increase of $\sim$\,0.37\,mm over the five years of operation.

This value is reasonably small compared to dead layer increases reported for similar detectors and time \cite{QUANGHUY2010390, loan_ge, Azbouche}, although a direct comparison is not possible given the different thermal cycles and crystal sizes. The newly measured thickness has been implemented in the Geant4 GeMSE model. Simulations of the commonly evaluated radioisotopes in low-background measurements indicate that the efficiency loss in the 100\,--\,2700\,keV range is not strongly affected by this increase, but becomes relevant for lines close to the energy threshold ($\sim$\,30\,keV) with branching ratios below the percent level.

\paragraph{Modeling Complex Geometries}

A common problem for the proper determination of the detection efficiency, and thus the sample activity, is the accurate implementation of the shape and placement of screened samples in the simulation. This is equally relevant for rare-event searches, for which counting rates are extremely low, and meteorite research, where the usual samples have complex, non-uniform shapes and therefore their $\upgamma$-ray self-absorption is not uniform either. The efficiencies for all meteorites screened with GeMSE so far~\cite{Rosen:2020aaa,Rosen:2021} were approximated by assuming the shape of an oblate ellipsoid.

To address this problem, the option to import 3D CAD models into Geant4 was added to the GeMSE simulation. The CADMesh interface~\cite{poole2012acad, poole2012fast} utilizes dedicated libraries to read the CAD file and directly converts it to a Geant4-readable format. The 3D CAD files can be generated using precision optical 3D scanners or even smartphone applications. The CAD models of the samples are imported as triangular facet surface meshes in binary format and returned as a \texttt{G4TessellatedSolid} object which can be handled in the Geant4 detector construction class.

As a complex-shaped example to showcase this feature, a banana was screened with GeMSE. Bananas are rich in potassium (0.36\% per mass~\cite{banana_potassium}), which includes \textsuperscript{40}K ($T$\textsubscript{1/2}\,=\,1.25\,$\times$\,$10^{9}$\,y) with a natural isotopic abundance of 0.0117\%. The \textsuperscript{40}K decay via electron capture, occurring in 11\% of the cases, leads to an \textsuperscript{40}Ar excited state which emits a 1461\,keV $\upgamma$-ray when de-exciting to the ground state. Given the high potassium content of a banana and the low-background regime of GeMSE, a measurement time of a few hours is already sufficient to identify this peak. However, its particularly complex shape is hard to implement with the default volume generators provided by Geant4.

\begin{figure}[t]
\centering
    \centering
    {\includegraphics[width=0.49\textwidth]{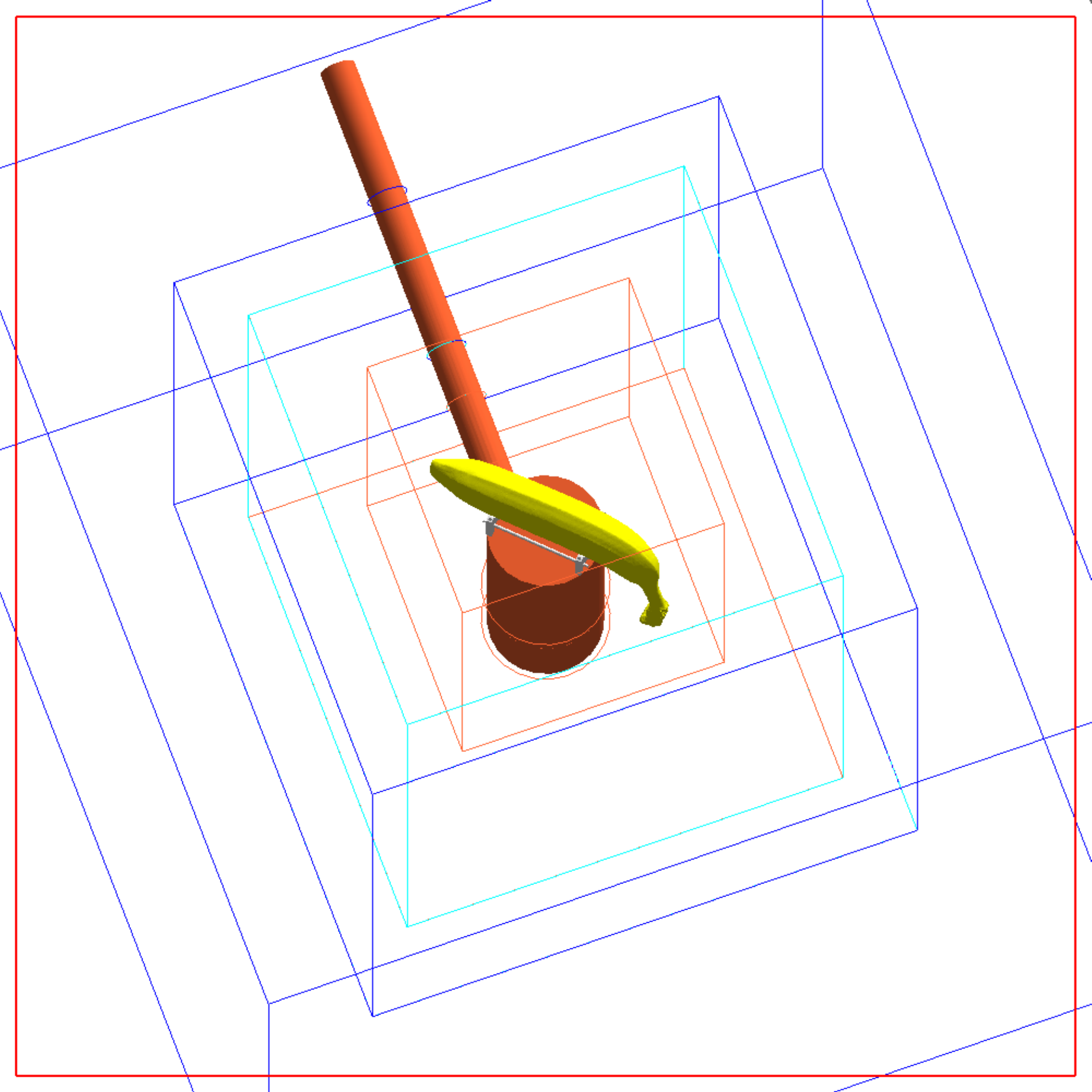}}
    {\includegraphics[width=0.49\textwidth]{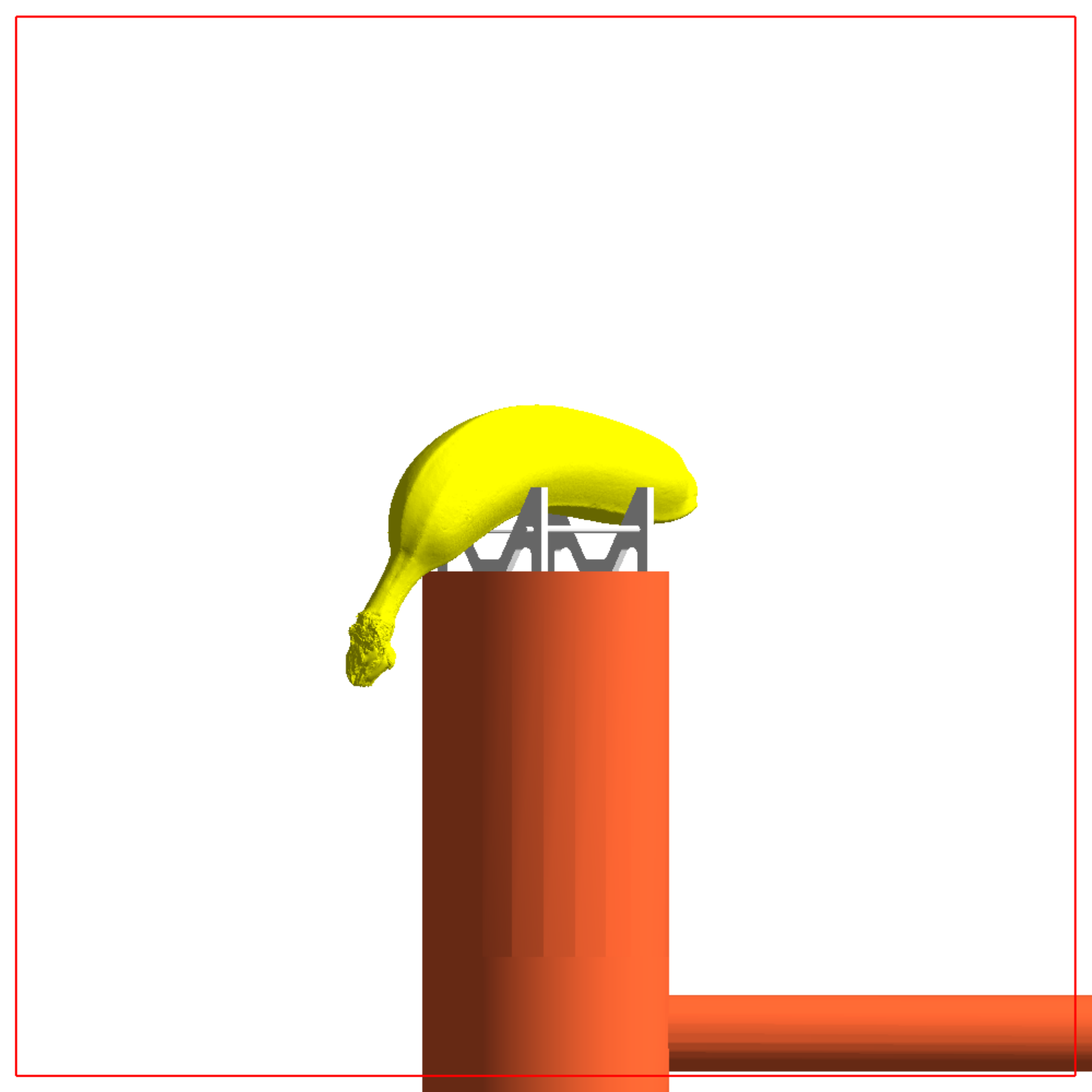}}
    \caption{\label{fig:g4_banana}Geant4 rendering of the screened banana inside the GeMSE sample cavity. The  sample was 3D-scanned and imported into the detector model. \textit{(Left)} The contours of the three-layer passive shielding: Cu-OFE (orange), low-activity Pb (cyan) and standard Pb (dark blue). The cold finger and the crystal's OFHC Cu cryostat (orange), the PTFE holder (white) and the banana are shown as solid bodies. \textit{(Right)} A close-up visualization of the precise implementation of the PTFE holder and the banana with respect to the cryostat endcap.}
\end{figure}

A high-resolution 3D scanner was used to generate the CAD model of the banana before placing it inside the detector cavity. For this purpose a light-weight PTFE sample holder was designed and also imported into the detector model. The simulated geometry is shown in figure~\ref{fig:g4_banana}. 10\textsuperscript{8} primary \textsuperscript{40}K isotope decays were generated within the simulated banana volume and the energy deposits in the HPGe crystal were recorded. The resulting detection efficiency of the Ge crystal for the full-absorption 1461\,keV $\upgamma$-rays is ($0.947\pm0.009$)\%, almost 10\% higher than the value obtained for a simplified cylinder placed on top of the PTFE sample holder, with similiar radius and length and equal volume and density as the banana.

\begin{SCfigure}
  \centering
  \captionsetup{format=plain, labelfont=bf}
  {\includegraphics[width=0.5\linewidth]{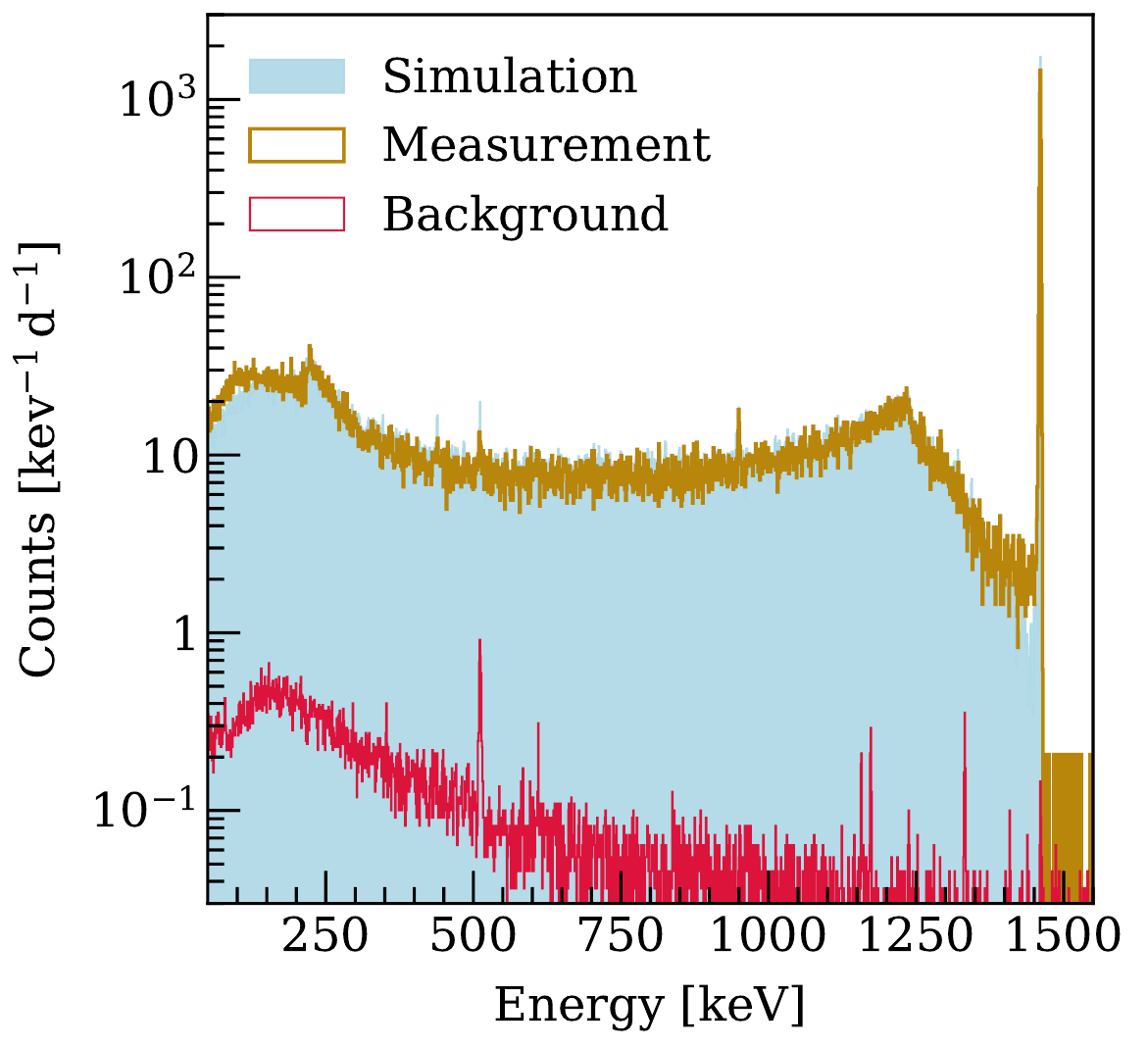}}
  \caption{Measured banana spectrum (brown) and GeMSE background (red, taken from figure~\ref{fig:gemse_bkg}). The simulated spectrum is also shown (blue) after a convolution with the detector resolution function. The backscatter peak at $\sim$\,250\,keV is visible, as well as the Compton edge at $\sim$\,1250\,keV. The area under the Gaussian full absorption peak at 1460.8\,keV matches that of the simulation.}
  \label{fig:banana_rate}
\end{SCfigure}

The spectrum acquired during a five-day run is shown in figure~\ref{fig:banana_rate} along with the simulated spectrum corrected for the detector energy resolution function. The measured and simulated area of the 1461\,keV peak match, although the measured spectrum features a tail towards lower energies and thus does not have the same height as the simulated $\upgamma$-peak. Events which survived the pile-up rejection of the data acquisition system or for which the charges were not entirely collected can be responsible for this. The impact of these artifacts is generally negligible as their contribution to the observed \textsuperscript{40}K peak is below the 0.1\% level. The Compton edge at $\sim$\,1250\,keV and the backscatter peak ($\sim$\,250\,keV) agree very well in data and simulation. The GeMSE background spectrum (red histogram from figure~\ref{fig:gemse_bkg}), several orders of magnitude lower than the banana spectrum, is also shown.

The \textsuperscript{40}K activity was calculated following the standard analysis approach described in~\cite{Sivers:2016yvm}, where the activity of the low-mass, low-background PTFE sample holder used for this measurement was neglected. The obtained activity is 42.6${+0.9 \atop -0.8}$\,Bq, which agrees with the usually quoted \textsuperscript{40}K activity for a standard banana. The mass proportion of potassium in banana pulp and skin are $\sim$\,0.36\%~\cite{banana_potassium} and $\sim$\,0.93\%~\cite{ARCHIBALD1949969}, respectively, and the banana skin contributes on average 36.6\% to the total mass~\cite{peel_mass}. Given the total banana mass of $m=265.15$\,g, this corresponds to 1.50\,g of 
\textsuperscript{nat}K in the fruit. The measured specific activity of \textsuperscript{nat}K is thus 42.6\,Bq\,/\,1.50\,g\,=\,(28.4\,$\pm$\,0.6)\,Bq/g, in excellent agreement with the literature values of 27.3\,--\,31.3\,Bq/g~\cite{Samat_1997}. This exercise demonstrates the potential improvement of screening measurements by simulating the precise three-dimensional shape of the screened sample. This is especially relevant for the radioassay of meteorite samples which in most cases must not be manipulated and often come in complex shapes.

%%%%%%%%%%%%%%%%%%%%%%%%%%%%%%%%%%%%%%%%%%%%%%%%%%%%%%%%%%%%%%

\section{Summary and Outlook}
\label{sec:summary}

The GeMSE low-background screening facility was installed in the Vue-des-Alpes underground laboratory with a moderate rock overburden in~2015 and has been operative since, with very limited downtime periods. Thanks to upgrades to the monitoring sensors, such as the newly developed LN\textsubscript{2} levelmeter for the dewar connected to the detector coldfinger, as well as to the custom-made slow control system Doberman~\cite{Doberman}, the facility installed in a rather remote location in a non-staffed laboratory is now automated and remotely controlled to a large extent. The autonomous operation time of the detector without the need for people on-site is now $\gtrsim$\,6\,weeks, larger than the typical sample measuring time.

\sloppy The most recent background run resulted in a rate of ($164\pm2$)\,counts/day or \mbox{($82\pm1$)\,counts/(day\,$\cdot\,$kg)} in the range 100\,--\,2700\,keV. The significant decrease of 33\% compared to the rate presented in~\cite{Sivers:2016yvm} after five years of underground operation agrees with the projected activity drop for the cosmogenically-activated isotopes. This has been verified and quantified by matching the experimental data with the simulated background contributions from a precise Geant4 model of the detector. GeMSE's background rate is now dominated by muon-induced events.

The spectrum of such muon-induced events in the HPGe crystal has been measured by using the muon veto to tag these events rather than discarding them. This allows extracting a muon veto efficiency of about 31\%, in good agreement with the Geant4 simulations of the detector, and confirms the shape of the muon-induced sprectrum used to fit the data. Further deactivation of the remaining short-lived isotopes in the Ge crystal and the shielding will only mildy reduce the background rate to 78\,counts/(day\,$\cdot\,$kg) in 2025, hence the next step is to target the dominating muon background: an additional top scintillator muon veto panel with larger geometrical coverage would achieve efficiencies of 50\,--\,60\%, reducing the background rate down to 55\,--\,40\,counts/(day\,$\cdot\,$kg).

The GeMSE detector model was improved to more accurately determine the detection efficiency for samples. The thickness of the HPGe crystal dead layer was investigated using the peak-ratio approach with a certified \textsuperscript{133}Ba calibration source and yielded a value of ($0.94\pm0.06$)\,mm, 0.37\,mm thicker than the measured value in 2016~\cite{Sivers:2016yvm}, probably caused by several thermal cycles over the past years. Furthermore, 3D geometries of complex-shaped objects can now be imported into the simulations model, thus mitigating the inaccuracies in the detection efficiency estimations for the examined samples. To exemplify this feature, a banana was screened with GeMSE and the detection efficiency was derived from the accurately implemented 3D model. This method will be highly advantageus for the future meteorite screening program.

%%%%%%%%%%%%%%%%%%%%%%%%%%%%%%%%%%%%%%%%%%%%%%%%%%%%%%%%%%%%%%

\section*{Acknowledgments}
This work was supported by the European Research Council (ERC) grant number~724320 (ULTIMATE) and the German  Ministry  for  Education and  Research (BMBF) grant number~05A17VF1. The facility was initially funded by the Swiss National Science Foundation (SNF) interdisciplinary grant number 152941. We gratefully acknowledge the continuous and remarkable support from the Albert Einstein Center for Fundamental Physics at the University of Bern, especially from Prof.~Michele Weber, Roger Hänni and his team, and Jean-Michel Villeumier. We thank Dr.~Rugard Dressler (PSI Villigen) for providing the calibration sources used to characterize the detector. We also thank the Bachelor students from the University of Freiburg who helped to improve GeMSE over the last years: Thomas Kok, Anne K.~Becker, Lukas B.~Grunwald and Baris Kiyim.

%%%%%%%%%%%%%%%%%%%%%%%%%%%%%%%%%%%%%%%%%%%%%%%%%%%%%%%%%%%%%%
%%%%%%%%%%%%%%%%%%%%%%%%%%%%%%%%%%%%%%%%%%%%%%%%%%%%%%%%%%%%%%
%%%%%%%%%%%%%%%%%%%%%%%%%%%%%%%%%%%%%%%%%%%%%%%%%%%%%%%%%%%%%%
\bibliography{main.bib}{}
\bibliographystyle{JHEP}

\end{document}